\documentclass[graybox]{svmult}

\usepackage{amsmath,amssymb}

\usepackage{txfonts}        
\usepackage{helvet}         
\usepackage{courier}        
\usepackage{type1cm}        
\usepackage{makeidx}         
\usepackage{multicol}        

\usepackage[bottom]{footmisc}

\usepackage{graphicx}        
\usepackage{subfigure}       

\usepackage{color}

\newcommand{\rd}{\mathrm{d}}

\def\doubleunderline#1{\underline{\underline{#1}}}

\usepackage[square,numbers]{natbib}

\usepackage{hyperref}

\makeindex       

\begin{document}

\title*{On the deformation of a hyperelastic tube due to steady viscous flow within}
\author{Vishal Anand and Ivan C. Christov}
\institute{Vishal Anand \at School of Mechanical Engineering, Purdue University, West Lafayette, Indiana 47907, USA \email{anand32@purdue.edu}
\and Ivan C. Christov \at School of Mechanical Engineering, Purdue University, West Lafayette, Indiana 47907, USA \email{christov@purdue.edu}}
\maketitle


\abstract{In this chapter, we analyze the steady-state microscale fluid--structure interaction (FSI) between a generalized Newtonian fluid and a hyperelastic tube. Physiological flows, especially in hemodynamics, serve as primary examples of such FSI phenomena. The small scale of the physical system renders the flow field, under the power-law rheological model, amenable to a closed-form solution using the lubrication approximation. On the other hand, negligible shear stresses on the walls of a long vessel allow the structure to be treated as a pressure vessel. The constitutive equation for the microtube is prescribed via the strain energy functional for an incompressible, isotropic Mooney--Rivlin material. We employ both the thin- and thick-walled formulations of the pressure vessel theory, and derive the static relation between the pressure load and the deformation of the structure. We harness the latter to determine the flow rate--pressure drop relationship for non-Newtonian flow in thin- and thick-walled soft hyperelastic microtubes. Through illustrative examples, we discuss how a hyperelastic tube supports the same pressure load as a linearly elastic tube with smaller deformation, thus requiring a higher pressure drop across itself to maintain a fixed flow rate.}

\section{Introduction}
\label{sec:intro}

Traditionally, physiological flows in soft and deformable tubes form a large class of literature on \emph{collapsible tubes} \cite{S77,GJ04,HH11}. These phenomena related to air flow in the lungs or blood flow in large blood vessels (such as arteries) are inherently a moderate-to-large-Reynolds number phenomenon. At the extreme of very large Reynolds number (inviscid) flow lies \emph{hydroelasticity} (to use the term of the group led by D.\ I.\ Indeitsev at the Institute of Problems of Mechanical Engineering Russian Academy of Sciences in St.\ Petersburg). Nowadays, there is a vast literature on hydroelasticity, covering both stable and unstable internal and external flows \cite{P98,GBGP18} that are capable of supporting nonlinear wave phenomena \cite{IO99}, including wave localization \cite{AIV98}. A common example of such high-Reynolds-number hydroelastic interactions is aerodynamic flutter \cite{BAH96}, which can lead to potentially disastrous instabilities such as failure of airplane wings and suspension bridges. A variety of nonlinear wave phenomena also arise when embedding a linearly elastic solid body (e.g., a rod) into an ambient viscoelastic medium and studying the coupled structure--structure interactions \cite{PV02}. More recently, \emph{however}, there has been significant interest in ``viscous--elastic'' fluid--structure interactions (FSIs) between internal fluid flows at low Reynolds numbers and soft tubes and annuli \cite{CM03,EG14,EG16}, including the effect of non-Newtonian rheology \cite{BBG17,RCDC18,AC18b}. This renewed interest comes from the need for understanding these systems in order to design microfluidic \cite{RCDC18} and soft robotic \cite{EG14} devices. At these smaller scales (or, for such ``creeping'' viscous flows), fluid inertia is negligible, and the balance between viscous fluid forces and elasticity is the dominant physical effect.

Here, we present a first foray into the mathematical analysis of low-Reynolds-number FSI, at steady state, due to the flow of a non-Newtonian fluid in a microtube composed of a \emph{hyperelastic} material. Hyperelastic materials have the ``advantage'' of being completely specified by a strain energy functional from which the constitutive equation between stress and strain follows. The structural response of complex soft solids, such as biological tissue and blood vessels, can be appropriately described by a hyperelastic solid with a \emph{pseudo} strain energy function (see, e.g., \cite[Ch.~8 and Ch.~9]{F93}). Similarly, due to its complex constituents, blood is a non-Newtonian fluid and an appropriate rheological model (beyond the simple Newtonian viscous fluid) should be used (see, e.g., \cite[Ch.~3]{F93}).

This chapter is thus organized as follows: in Sec.~\ref{section:flow}, we address the fluid mechanics problem, including the velocity profile of a generalized Newtonian (specifically, power-law) fluid in a tube of slowly varying cross-section. In Sec.~\ref{section:structure}, we discuss the deformation of the soft hyperelastic microtube due to uniform loading from within. Then, in Sec.~\ref{section:coupling}, we specifically choose the load to be the hydrodynamic pressure and obtain the appropriate pressure--deformation relations describing such fluid--structure interactions. In Sec.~\ref{sec:result}, the results are discussed and compared to limiting cases, such as a linearly elastic microtube, in order to highlight the effects of hyperelasticity. Conclusions and avenues for future work are stated in Sec.~\ref{sec:conclusion}. An appendix is included for completeness, in which the pressure--deformation and flow rate--pressure drop relations for FSI in a thick-walled linearly elastic tube are also derived.

\section{Summary of the fluid mechanics problem}
\label{section:flow}
\begin{figure}
    \centering
    \includegraphics[width=0.75\textwidth]{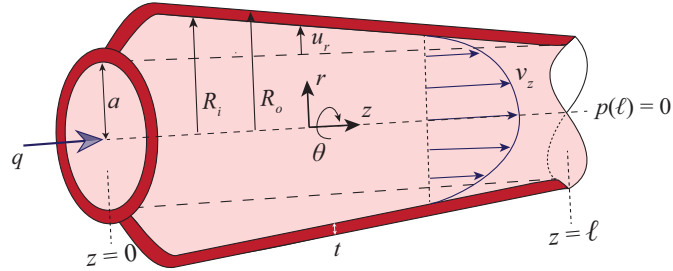}
    \caption{Schematic of the microtube geometry, including notation used.}
    \label{fig:schematic}
\end{figure}

Consider a fluid flow $\vec{v} = v_r \hat{\vec{r}} + v_\theta \hat{\vec{\theta}} + v_z \hat{\vec{z}}$ in cylindrical coordinates. A diagram of the deformed microtube geometry is shown in Fig.~\ref{fig:schematic}, specifically the tube has uniform thickness $t$, undeformed inner radius $a$, and length $\ell$. Now, following \cite{AC18b}, let us assume that
\begin{enumerate}
\item The flow is steady: $\frac{\partial}{\partial t}(\,\cdot\,) = 0$.

\item The flow is axisymmetric: $\frac{\partial}{\partial \theta}(\,\cdot\,) = 0$ and $v_{\theta} = 0$.

\item The geometry of the flow vessel is a slender tube: $\ell \gg a \;\Leftrightarrow\; {a}/{\ell} = \epsilon \ll 1$.
\end{enumerate}
Assumption 3.\ is key to the mathematical analysis below. Specifically, this assumption leads us to the appeal to the \emph{lubrication approximation} of fluid mechanics (see, e.g., \cite{panton,L07}), which will allow us to solve for the flow profile analytically.

As shown with due diligence in our previous work \cite{AC18b}, to the leading order in $a/\ell$, the velocity field is \emph{unidirectional} \cite{L07}: $\vec{\bar{v}} = \bar{v}_{\bar{z}} \, \hat{\vec{z}}$ but weakly varying with the (long) flow-wise direction i.e., $\bar{v}_{\bar{z}}=\bar{v}_{\bar{z}}(\bar{r},\bar{z})$. Here, and henceforth, bars over quantities denote dimensionless variables according to the following scheme:
\begin{equation}
   \bar{r} = r/a, \qquad
   \bar{z} = z/\ell, \qquad
   \bar{v}_{\bar{r}} = v_r/\mathcal{V}_r, \qquad   
   \bar{v}_{\bar{z}} = v_z/\mathcal{V}_z, \qquad
   \bar{p} = p/\mathcal{P}_c, 
\label{eq:nd_vars_tube}
\end{equation}
where the characteristic radial velocity scale is $\mathcal{V}_r \equiv  \epsilon \mathcal{V}_z$, the characteristic axial velocity scale is $\mathcal{V}_z$, and $\mathcal{P}_c$ is the characteristic pressure scale. Upon specifying the rheology of the fluid (see below), $\mathcal{P}_c$ and $\mathcal{V}_z$ will be related to each other.
 
Next, we specify the fluid's rheological behavior. We are interested in biofluid mechanics applications such as blood flow through a deformable artery or vein. Blood is known to have a shear-dependent viscosity due to the fact that red blood cells deform. Haemorheology is a complex topic \cite[Ch.~3]{F93}, nevertheless experiments suggest \cite{HKP99} that blood flow can be accurately fit to a power-law fluid model (a generalized Newtonian rheology often going by the name Ostwald--de Waele \cite{Bird76}) at steady state. Now, the dominant shear stress component is $\tau_{rz}$; likewise the corresponding rate-of-strain tensor component is just $\dot{\gamma}_{rz}=\partial v_z /\partial r$ to the leading order in $a/\ell$ (i.e., under the slenderness assumption). Thus the fluid's rheological model takes the ``simple shear'' form:
\begin{equation}
\label{stress_mt}
   \tau_{rz} = \underbrace{m\left|\frac{\partial v_z}{\partial r}\right|^{n-1}}_{=\eta} \frac{\partial v_z}{\partial r},
\end{equation}
where $\eta$ is the \emph{apparent viscosity}, $m(>0)$ is the \emph{consistency factor}, and $n(>0)$ is the \emph{power-law index}. On making Eq.~\eqref{stress_mt} dimensionless using the variables from Eq.~\eqref{eq:nd_vars_tube}, we obtain the relationship between the axial velocity $\mathcal{V}_z$ scale and the pressure scale $\mathcal{P}_c$: $\mathcal{V}_z = \left[{a^{n+1}\mathcal{P}_c}/{(m\ell)}\right]^{1/n}$. 

The power-law rheological model captures the flow behavior of fluids that are shear thinning ($\eta$ decreases with $\dot{\gamma}_{rz}$), such as blood \cite{Chien66,HKP99} as mentioned above, for $n<1$. It also captures the flow behavior of shear-thickening fluids ($\eta$ increases with $\dot{\gamma}_{rz}$) such as as woven Kevlar fabrics impregnated with a colloidal suspension of silica particles \cite{WB09} used for ballistic armors, for $n>1$. The viscous Newtonian fluid is obtained as the special case of $n=1$ in Eq.~\eqref{stress_mt}. Our motivation is mainly fluid mechanics of blood vessels, so our examples consider $n<1$, but the theory applies to both cases.

Finally, considering the dynamics of the flow under the constitutive relationship in Eq.~\eqref{stress_mt}, we found that the dimensionless axial velocity profile is \cite{AC18b}:
\begin{equation}
\label{eq:barVz_tube}
 \bar{v}_{\bar{z}} = \left(-\frac{1}{2}\frac{\rd \bar{p}}{\rd \bar{z}}\right)^{1/n}\left(\frac{\bar{R}_i^{1+1/n}-\bar{r}^{1+1/n}}{1+1/n}\right),
\end{equation}
where $\bar{R}_i = R_i/a$ is the dimensionless \emph{inner} radius of the deformed microtube. More importantly, $\bar{R}_i$ is not necessarily equal to unity because we allow the microtube to deform due to FSI, as discussed in the next section. Likewise, the pressure gradient $\rd \bar{p}/\rd \bar{z}$ is \emph{not} constant but, rather, varies with $\bar{z}$. As a result, while $\bar{p}$ is at most a function of $\bar{z}$ (but not a linear function), $\bar{v}_{\bar{z}}$ can depend upon both $\bar{r}$ and $\bar{z}$.

\section{Structural mechanics problem: solving for the deformation}
\label{section:structure}

In this section, we address the structural mechanics aspect of the coupled FSI problem posed above. To this end, we treat the structure as a \emph{pressure vessel}, wherein the only load acting on the structure is the hydrodynamic pressure from the fluid, and the load due to viscous and shear stresses is neglected. This assumption stems from the lubrication approximation for the fluid flow, wherein the viscous shear stress scale is $\approx \epsilon$ times the hydrodynamic pressure scale \cite[Ch.~22]{panton}. We begin our discussion by first analyzing a \emph{thin-walled} pressure vessel, then we move onto its \emph{thick-walled} counterpart.

\subsection{Thin-walled cylinder}

Let us consider the case of a thin-walled initially cylindrical microtube with thickness $t \ll a$. This assumption allows us to consider the cylinder in a state of plane stress and plane strain, thus simplifying the analysis of the structural mechanics problem. As a consequence, the walls of the cylinder act like a membrane, which does not sustain any bending or twisting moments. There is no variation of stress and deformation throughout the thickness of the cylinder.

\subsubsection{Deformation}
 \label{Sec:Kinematics_Thin}

In the undeformed material coordinate system, the coordinates of a material point are given by $r =a $, $\theta \in [0,2\pi]$, and $z \in [0,\ell]$. Upon \emph{axisymmetric} deformation, the coordinates of the same material point become
\begin{equation}
     R = R(r), \qquad \Theta = \theta, \qquad Z = \alpha z.
\end{equation}
We further assume that the deformation is homogeneous along the axial direction, thus $\alpha =L/\ell$, with $L$ being the deformed cylinder's length and $\ell$ being the undeformed cylinder's length (as in Fig.~\ref{fig:schematic}). Now, since the cylinder is clamped at both its ends, its length does not change and $L=\ell$. Hence, $\alpha = 1$. The foregoing discussion reduces the coordinates of the point in the spatial coordinate system to:
\begin{equation}
     R = R(r), \qquad \Theta = \theta, \qquad Z = z.
\label{eq:Deformed_Coordinates}
\end{equation}
As the shell is considered (infinitesimally) thin in this theory, we denote the inner radius $R_i$ by $R$ (in this section) without fear of confusion.

For the case of a deformation defined by Eq.~\eqref{eq:Deformed_Coordinates}, the deformation gradient tensor $\doubleunderline{\mathcal{F}}$ can be easily computed:
\begin{equation}
\label{eq:deformation_tensor_reduced}
\doubleunderline{\mathcal{F}} = 
\begin{pmatrix}
{\partial R}/{\partial r} & 0 & 0 \\
0&  {R}/{r}& 0\\
0 & 0 & 1
\end{pmatrix}.
\end{equation}
Since $\doubleunderline{\mathcal{F}}$ is a diagonal tensor, then its principal axes are just the $r$, $\theta$, and $z$ axes of the cylindrical coordinate system. Indeed, we deduce from the deformation field introduced in Eq.~\eqref{eq:Deformed_Coordinates} that a line segment oriented along either the $r$, $\theta$ or $z$ coordinate directions will, at most, only \emph{stretch} and cannot rotate.

Consequently, for this type of deformation, the rotation tensor $\doubleunderline{\mathcal{R}} = \doubleunderline{\mathcal{I}} $ (the identity tensor) and  the stretch tensor is simply
\begin{equation}
\label{eq:stretch_tensor}
\doubleunderline{\mathcal{U}} = \doubleunderline{\mathcal{F}} = \begin{pmatrix}
\lambda_1 & 0 & 0 \\
0&  \lambda_2& 0\\
0 & 0 & \lambda_3 
\end{pmatrix}.
\end{equation}
Here, $\lambda_1$, $\lambda_2$, $\lambda_3$ are the principal stretches, which one can immediately write down by comparison of Eqs.~\eqref{eq:deformation_tensor_reduced} and \eqref{eq:stretch_tensor}. Now, since the material is incompressible, $\det \doubleunderline{\mathcal{F}} = \lambda_1\lambda_2\lambda_3=1$, we can determine $\lambda_1$, and thus
\begin{equation}
    \lambda_1 = r/R, \qquad \lambda_2 = R/r, \qquad \lambda_3 = 1.
\label{eq:principal_stretches}
\end{equation}

\subsubsection{Constitutive equation}
\label{Sec:Constitutive_Thin}
We consider the material from which the cylindrical tube is composed to be an isotropic, incompressible, \emph{hyperelastic} material. For such a material, the constitutive equation is specified through a strain energy functional $W$ \cite{BW08,BW00}, which depends upon the principal stretches $\lambda_i$, i.e., $W=W(\lambda_1, \lambda_2, \lambda_3)$. Specifically, we assume that the hyperelastic material is defined by the \emph{incompressible} Mooney--Rivlin constitutive equation \cite{L05,BW08,BW00} with strain energy given by
\begin{equation}
    W = \mathbb{C}_1 \left(\lambda_1^2+\lambda_2^2+\lambda_3^2\right) + \mathbb{C}_2 \left(\lambda_1^2\lambda_2^2 + \lambda_2^2\lambda_3^2+\lambda_3^2\lambda_1^2\right) \qquad(\lambda_1\lambda_2\lambda_3 =1).
    \label{eq:constitutive_equation}
\end{equation}
Here, $\mathbb{C}_1$ and $\mathbb{C}_2$ are two material constants characterizing the structural response of the hyperelasic material; they are determined empirically by comparison to experiments \cite{O72}. Equation~\eqref{eq:constitutive_equation} is traditionally invoked to describe the response of highly elastic, i.e., ``rubber-like,'' materials under isothermal conditions \cite{O72}. In particular, setting $\mathbb{C}_2 = 0$ reduces the Mooney--Rivlin model to the neo-Hookean solid. For most ``rubber-like'' materials, $\mathbb{C}_1 >0$ and $\mathbb{C}_2 \le 0$ \cite{Liu2012, Moon1974, Mihai2011}. For compatibility with linear (i.e., small-strain) elasticity theory (see \cite[Eq.~(6.11.29)]{BW00}), we must have
\begin{equation}
    G = 2(\mathbb{C}_2+\mathbb{C}_1)
    \label{eq:Linear_Hyper}
\end{equation}
as the shear modulus of elasticity. We also recall that, for a linearly elastic material,
\begin{equation}
    2G(1+\nu) = E, 
    \label{eq:linear-elastic}
\end{equation}
where $\nu$ is the Poisson ratio, and $E$ is Young's modulus.

Now, for isotropic materials, the principal Cauchy stresses are coaxial with the principal stretches and are given by
\begin{subequations}\begin{align}
    \sigma_1 - \sigma_3 &= \lambda_1\frac{\partial W}{\partial \lambda_1}-\lambda_3\frac{\partial W}{\partial \lambda_3}, \\
    \sigma_2 - \sigma_3  &= \lambda_2\frac{\partial W}{\partial \lambda_2}-\lambda_3\frac{\partial W}{\partial \lambda_3}.
\end{align}\label{eq:StressEquation}\end{subequations}
Substituting Eq.~\eqref{eq:constitutive_equation} into Eqs.~\eqref{eq:StressEquation}, we obtain
\begin{subequations}\label{eq:constitutive_sigmas}\begin{align}
    \sigma_1 - \sigma_3 &= 2\mathbb{C}_1\left(\lambda_1^2-\lambda_3^2\right)-2\mathbb{C}_2\left(\lambda_1^{-2}-\lambda_3^2\right), \\
    \sigma_2 -\sigma_3 &= 2\mathbb{C}_1\left(\lambda_2^2-\lambda_3^2\right)-2\mathbb{C}_2\left(\lambda_2^{-2}-\lambda_3^2\right).
\end{align}\end{subequations}

\subsubsection{Static equilibrium}
\label{eq:Static_Thin}

As mentioned above, our exemplar thin-walled cylinder acts as a {pressure vessel}, i.e., a structure that sustains only stretching and tension but no bending or twisting. For such a structure, the equations of static equilibrium take the form:
\begin{subequations}\begin{align}
    \sigma_{1} &= \sigma_{rr}= -p, \\
    \sigma_{2} &= \sigma_{\theta\theta} = \frac{pR}{t},\\
    \sigma_{3} &= \sigma_{zz}= \frac{pR}{2t}.
\end{align}\label{eq:sigma_11_22_33}\end{subequations}
Since the tube is thin, i.e., $t\ll a$ and $R=\mathcal{O}(a)$, then
$\sigma_{1}\ll \sigma_{2}\approx \sigma_{3}$. Substituting Eqs.~\eqref{eq:sigma_11_22_33} (stress balance) into Eqs.~\eqref{eq:constitutive_sigmas} (constitutive) and employing Eqs.~\eqref{eq:principal_stretches} (deformation), we obtain 
\begin{subequations}\begin{align}
    -\frac{pR}{2t} &= 2\mathbb{C}_1\left(\frac{a^2}{R^2}-1\right)-2\mathbb{C}_2\left(\frac{R^2}{a^2}-1\right),\displaybreak[3]\\
    \frac{pR}{2t} &= 2\mathbb{C}_1\left(\frac{R^2}{a^2}-1\right)-2\mathbb{C}_2\left(\frac{a^2}{R^2}-1\right).
\end{align}\end{subequations}
Combining the last two equations, we arrive at the pressure--radius relation
\begin{equation}
    \frac{pa}{2t(\mathbb{C}_1+\mathbb{C}_2)} = \frac{R}{a}-\frac{a^3}{R^3},
    \label{eq:Deformation_Dim_Final}
\end{equation}
where $R = a + u_r$ is the deformed tube radius, and $u_r$ is the radial deformation (recall Fig.~\ref{fig:schematic}). 

Notice that the cross-sectional area of the tube at some fixed axial location, $z$, is $A = \pi R^2$ [here, $R=R(z)$ due to FSI]. Then, Eq.~\eqref{eq:Deformation_Dim_Final} can be rewritten as a \emph{pressure--area} relationship:
\begin{equation}
    p(\bar{A}) =  2(\mathbb{C}_1+\mathbb{C}_2)\frac{\bar{t}}{\sqrt{\bar{A}}}\left(\bar{A}-\frac{1}{\bar{A}}\right),
    \label{eq:tube_law_thin}
\end{equation}
where $\bar{t}=t/a$ is the dimensionless (reduced) thickness of the tube, and $\bar{A} = A/(\pi a^2)$ is the dimensionless (reduced) area of the deformed tube under axisymmetric conditions (initial circular cross-section remains circular under deformation). Equation~\eqref{eq:tube_law_thin} represent a \emph{tube law} \cite{WHJW10} for microscale FSI in a hyperelastic pressure vessel. This ``law'' is often used as a ``constitutive'' equation (closure) in unsteady FSI problems in which the flow is cross-sectionally averaged \cite{S77}. Of interest is to note that $p(\bar{A})$ in Eq.~\eqref{eq:tube_law_thin} is \emph{nonlinear}.

Finally, let us make Eq.~\eqref{eq:Deformation_Dim_Final} dimensionless using the following dimensionless variables
\begin{equation}
    \bar{u}_{\bar{r}} = u_r/a, \qquad \bar{p} = p/\mathcal{P}_c,
    \label{eq:dimless_vars_structure}
\end{equation}
to yield 
\begin{equation}
    \gamma\bar{p} = (1+\bar{u}_{\bar{r}})-\frac{1}{(1+\bar{u}_{\bar{r}})^3},\qquad \gamma := \frac{\mathcal{P}_c}{2(\mathbb{C}_1+\mathbb{C}_2)\bar{t}},
    \label{eq:pressure_deformation_dimless_thin}
\end{equation}
where $\bar{t}=t/a$ as above, and we have defined $\gamma$ as a dimensionless parameter that captures the ``strength'' of fluid--structure coupling, i.e., the so-called \emph{FSI parameter}. In our previous work on linearly elastic incompressible microtubes \cite{AC18b}, the FSI parameter was defined as $\beta = \mathcal{P}_c/(E\bar{t})$. To connect the hyperelastic theory to the linearly elastic theory, we can use Eqs.~\eqref{eq:Linear_Hyper} and \eqref{eq:linear-elastic}, taking $\nu = 1/2$ for an incompressible material, to find that
\begin{equation}
    \gamma = \beta/3.
    \label{eq:beta_gamma}
\end{equation}

A few remarks are in order. Equation~\eqref{eq:pressure_deformation_dimless_thin} represents the final dimensionless form of the pressure--deformation relation for a thin-walled incompressible hyperelastic cylinder. Second, note that, being a quartic (polynomial) equation in $(1+\bar{u}_{\bar{r}})$, Eq.~\eqref{eq:pressure_deformation_dimless_thin} can be solved explicitly for $\bar{u}_{\bar{r}}$ as a function of $\bar{p}$ using, e.g., {\sc Mathematica}. However, the resulting expression is too lengthy to be worth including here. Third, observe that both $\bar{p}$ and $\bar{u}_{\bar{r}}$ can (and do) depend on the dimensionless flow-wise coordinate $\bar{z}$, but $\bar{z}$ does not feature explicitly in the pressure--deformation relationship in Eq.~\eqref{eq:pressure_deformation_dimless_thin}.

\subsection{Thick-walled cylinder}

In this section, we account for the non-negligible thickness of a cylinder, i.e., the case of thick-walled pressure vessel, also known as \emph{Lam\'e's first problem} (see, e.g., \cite{M77}).

Unlike the case of a thin-walled cylinder, the inner and outer radii of the thick-walled cylinder differ. They are, thus, denoted by $r_i$ and $r_o$ before deformation, and by $R_i$ and $R_o$ after deformation. Specifically,
\begin{subequations}\begin{align}
r_i &= a, \\
r_o &= a + t,\\
R_i &= a + u_r,
\end{align}\label{eq:Geometry_Inner_Outer}\end{subequations}
where $u_r$ is the radial displacement of the inner surface of the cylinder. For the problem that we have posed, the displacement of the outer surface is of no consequence to the flow within the cylinder, hence we do not discuss it; then, denoting the displacement of the inner surface by $u_r$ is unambiguous. Since the cylinder's wall is assumed to be composed of an incompressible material (constant volume), and it is clamped at both its ends (constant length), its cross sectional area remains constant. Therefore,
\begin{equation}
\label{eq:Area_Constant}
R_o^2-R_i^2= r_o^2-r_i^2.
\end{equation}

The cylinder kinematics (Sec.~\ref{Sec:Kinematics_Thin}) and the hyperelastic constitutive equations (Sec.~\ref{Sec:Constitutive_Thin}) developed for the thin-walled cylinder also apply to the thick-walled one. However, unlike the previously discussed case, the stress and deformation fields are not constant for a thick-walled cylinder. Specifically, the stress and deformation vary across the thickness of the cylinder. Consequently, the equations for static equilibrium of a thick-walled tube are differential equations. Neglecting body forces (due to the small scale of the posed FSI problem) and shear stresses, the equations for static equilibrium \cite{L05,D90,Kraus67,Flugge60} are thus:
\begin{subequations}\begin{align}
\frac{\partial \sigma_{rr}}{\partial R}+\frac{1}{R}(\sigma_{rr}-\sigma_{\theta\theta}) &= 0, \\
\frac{\partial \sigma_{zz}}{\partial Z} &= 0.
\end{align}\end{subequations}
The latter equations, when written in terms of material coordinates in association with the deformation field described in Eq.~\eqref{eq:Deformed_Coordinates}, reduce to
\begin{subequations}\begin{align}
\frac{\partial (R\sigma_{rr})}{\partial r}+\frac{r}{R}\sigma_{\theta\theta} &= 0, \label{eq:stress_balance_r}\\
\frac{\partial \sigma_{zz}}{\partial z} &= 0. \label{eq:stress_balance_z}
\end{align}\label{eq:stress_balance}\end{subequations}
Note that Eq.~\eqref{eq:stress_balance_z} is satisfied identically.

As above, $\sigma_{rr}(=\sigma_{1})$ and $\sigma_{\theta \theta } (=\sigma_{2})$ are also related by Eq.~\eqref{eq:constitutive_sigmas}, i.e., the principal stress relations for an isotropic, hyperelastic Mooney--Rivlin material. Thus, we can eliminate $\sigma_{\theta\theta}$ from the constitutive equation~\eqref{eq:constitutive_sigmas} and the static equilibrium equations~\eqref{eq:stress_balance}. Then, employing the expressions for the principal stretches $\lambda_1$ and $\lambda_2$ from Eq.~\eqref{eq:principal_stretches}, we obtain the following differential equation governing $\sigma_{rr}$:
\begin{equation}
    \frac{\partial(R \sigma_{rr})}{\partial r}+\frac{r}{R}\left\{\sigma_{rr}+2(\mathbb{C}_1+\mathbb{C}_2)\left [\left(\frac{R}{r}\right)^2-\left(\frac{r}{R}\right)^2\right]\right\}=0.
    \label{eq:sigma_rr_ode}
\end{equation}
We solve Eq.~\eqref{eq:sigma_rr_ode} for $\sigma_{rr}$ subject to the loading boundary conditions
\begin{subequations}\begin{align}
    \sigma_{rr}|_{r = r_o} &= 0, \\
    \sigma_{rr}|_{r = r_i} &= -p,
\end{align}\end{subequations}   
to obtain the pressure--radius relation
\begin{equation}
\label{eq:pressure_deformation_thick}
p = \left[f\left(\frac{r_1^2}{R_1^2}\right)-f\left(\frac{r_i^2}{R_i^2}\right)\right](\mathbb{C}_1+\mathbb{C}_2),
\end{equation}
where, for convenience, the function $f$ is defined (see also \cite{L05}) as
\begin{equation}
f(\xi) := \xi +\ln \xi.
\label{eq:fx}
\end{equation}
Note that  Eq.~\eqref{eq:pressure_deformation_thick} is equivalent to Eq.~(3.4.3) in \cite[Ch.~9]{L05}. It is relevant to remind the reader that $p\ne p(r)$, so the integration of Eq.~\eqref{eq:sigma_rr_ode} is straightforward.

Finally, substituting the geometric relationships from Eqs.~\eqref{eq:Geometry_Inner_Outer} and ~\eqref{eq:Area_Constant} into Eq.~\eqref{eq:pressure_deformation_thick}, we obtain
\begin{equation}
 \frac{p}{\mathbb{C}_1+\mathbb{C}_2} = \left[f\left(\frac{(1+t/a)^2}{(1+t/a)^2-1+(1+u_r/a)^2}\right)-f\left(\frac{1}{(1+u_r/a)^2}\right)\right].
\end{equation}
The last equation can be re-written in dimensionless form using the variables from Eq.~\eqref{eq:dimless_vars_structure} as
\begin{equation}
     2\bar{t}{\gamma}{\bar{p}} = \left[f\left(\frac{(1+\bar{t})^2}{(1+\bar{t})^2-1+(1+\bar{u}_{\bar{r}})^2}\right)-f\left(\frac{1}{(1+\bar{u}_{\bar{r}})^2}\right)\right], 
     \label{eq:deflection_pressure_thick_nondim}
\end{equation}
where ${\gamma}$ is the FSI parameter as defined in Eq.~\eqref{eq:pressure_deformation_dimless_thin} above. Equation~\eqref{eq:deflection_pressure_thick_nondim} represents the final dimensionless form of the pressure--deformation relation for a thick-walled incompressible hyperelastic cylinder.

\section{Coupling of the fluid and structural mechanics problems}
\label{section:coupling}

We now turn to the main task, which is coupling the flow and deformation. As shown in our previous work \cite{AC18b}, this task is accomplished by computing the flow rate $q$ explicitly using its definition for an axisymmetric cylindrical tube:
\begin{equation}
q = \int_{0}^{2\pi}\int_{0}^{R_i}v_z \, r \,\rd r\, \rd\theta 
= \mathcal{V}_z 2\pi a^2 \int_{0}^{{\bar{R}_i}}{\bar{v}}_{\bar{z}} \, \bar{r} \,\rd\bar{r},
\label{eq:q_defn}
\end{equation}
where we have also introduced the dimensionless variables from Eq.~\eqref{eq:nd_vars_tube}. Now, substituting the expression for 
$\bar{v}_{\bar{z}}$ from Eq.~\eqref{eq:barVz_tube} into Eq.~\eqref{eq:q_defn}, we obtain the \emph{dimensionless} flow rate
\begin{equation}
\label{recipe_for_ODE_numerical}
\bar{q} \equiv \frac{q}{\mathcal{V}_z \pi a^2} =  \left(-\frac{1}{2}\frac{\rd\bar{p}}{\rd\bar{z}}\right)^{1/n}\frac{\bar{R}_i^{3+1/n}}{3+1/n},
\end{equation}
where $\bar{R}_i=1+\bar{u}_{\bar{r}}$ is  the dimensionless \emph{inner} radius of the deformed tube.

Thus, since $\bar{q}=const.$ by conservation of mass in a steady flow \cite{panton}, Eq.~\eqref{recipe_for_ODE_numerical} becomes an ordinary differential equation (ODE) for $\bar{p}(\bar{z})$:
\begin{equation}
\frac{\rd \bar{p}}{\rd \bar{z}} = -2[(3+1/n)\bar{q}]^n[1+\bar{u}_{\bar{r}}(\bar{z})]^{-(1+3n)}.
\label{eq:dpdz_leading_order}
\end{equation}
Now, we must specify the deformation profile  $\bar{u}_{\bar{r}}$ to complete the calculation. Even in the special case of a Newtonian fluid ($n=1$), Eq.~\eqref{eq:dpdz_leading_order} represents a strongly nonlinear pressure gradient--deformation coupling.

\subsection{Thin-walled cylinder}
\label{sec:coupling_thin}

To finish the derivation of the coupled FSI theory for a thin-walled cylinder, we differentiate the pressure--deformation relation from Eq.~\eqref{eq:pressure_deformation_dimless_thin} with respect to $\bar{z}$ to obtain
\begin{equation}
    \gamma\frac{\rd \bar{p}}{\rd \bar{z}} = \left[1 + \frac{3}{(1+\bar{u}_{\bar{r}})^4}\right] \frac{\rd \bar{u}_{\bar{r}}}{\rd \bar{z}}.
    \label{eq:dp_dz_deformation_dimless}
\end{equation}
Then, we eliminate $\rd \bar{p}/\rd \bar{z}$ between Eqs.~\eqref{eq:dpdz_leading_order} and \eqref{eq:dp_dz_deformation_dimless} to obtain an ODE for $\bar{u}_{\bar{r}}$:
\begin{equation}
  -2\gamma[(3+1/n)\bar{q}]^n=  \left[(1+\bar{u}_{\bar{r}})^{3n+1}+(1+\bar{u}_{\bar{r}})^{3n-3}\right]\frac{\rd \bar{u}_{\bar{r}}}{\rd \bar{z}}.
\label{eq:deflection_ODE_nondim_thin}
\end{equation}
Since conservation of mass dictates that $\bar{q}=const.$, and $\gamma$ and $n$ are known constants, the ODE~\eqref{eq:deflection_ODE_nondim_thin} can be separated and directly integrated, subject to the boundary condition (BC)  $\bar{u}_{\bar{r}}(\bar{z}=1)=0$,\footnote{Note, more importantly, that although in general we cannot expect to satisfy \emph{clamping} BCs, i.e., $\bar{u}_{\bar{r}}=\rd \bar{u}_{\bar{r}}/\rd \bar{z}=0$ at $\bar{z}=0$ and $\bar{z}=1$ in this leading-order analysis of deformation, we \emph{must} respect the pressure outlet BC, i.e., $\bar{p}(\bar{z}=1)=0$. From Eq.~\eqref{eq:pressure_deformation_dimless_thin}, it is then clear that the pressure BC requires that $\bar{u}_{\bar{r}}(\bar{z}=1)=0$ as assumed.} to yield:
\begin{equation}
 2\gamma[(3+1/n)\bar{q}]^n(1-\bar{z}) =  \frac{[1+\bar{u}_{\bar{r}}(\bar{z})]^{3n+2}}{(3n+2)}+\frac{[1+\bar{u}_{\bar{r}}(\bar{z})]^{3n-2}}{(3n-2)} - \frac{6n}{(3n+2)(3n-2)}.
\label{eq:flowrate_deflection_nondim}
\end{equation}
 
Equations~\eqref{eq:pressure_deformation_dimless_thin} and \eqref{eq:flowrate_deflection_nondim} fully specify (albeit implicitly) the static response of a thin hyperelastic cylinder due to  internal flow of a generalized Newtonian fluid within it. For example, the displacement at $\bar{z}=0$ found from Eq.~\eqref{eq:flowrate_deflection_nondim} can be used in Eq.~\eqref{eq:pressure_deformation_dimless_thin} to determine $\bar{p}(0)$ from which the full dimensionless pressure drop follows: $\Delta\bar{p} := \bar{p}(0) - \bar{p}(1)$, where $\bar{p}(1)=0$ is our chosen pressure gauge for the pressure at the outlet and also in the surrounding medium exterior to the cylinder. Thus, the flow rate--pressure drop relationship ($\bar{q}$ as a function of $\Delta\bar{p}$, or vice versa), i.e., a generalized Hagen--Poiseuille law, in the presence of FSI can be obtained analytically.

\subsection{Thick-walled cylinder}
\label{sec:coupling_thick}

For a thick-walled cylinder, we differentiate Eq.~\eqref{eq:deflection_pressure_thick_nondim} with respect to $\bar{z}$ to obtain
\begin{multline}
     {\gamma} \frac{\rd \bar{p}}{\rd \bar{z}} = -\left\{\frac{2(1+\bar{t})^2(1+\bar{u}_{\bar{r}})}{[(1+\bar{t})^2+(\bar{u}_{\bar{r}})^2+2\bar{u}_{\bar{r}}]^2}+\frac{2(1+\bar{u}_{\bar{r}})}{(1+\bar{t})^2+(\bar{u}_{\bar{r}})^2+2\bar{u}_{\bar{r}}} \right. \\
    \left. -\frac{2}{(1+\bar{u}_{\bar{r}})}-\frac{2}{(1+\bar{u}_{\bar{r}})^3}\right\} \frac{\rd\bar{u}_{\bar{r}}}{\rd\bar{z}}.
\label{eq:deflection_pressure_gradient_thick_nondim}
\end{multline}
Then, we eliminate $\rd \bar{p}/\rd \bar{z}$ between Eqs.~\eqref{eq:dpdz_leading_order} and \eqref{eq:deflection_pressure_gradient_thick_nondim} to obtain an ODE for the dimensionless transverse deflection $\bar{u}_{\bar{r}}$:
\begin{multline}
\left\{\frac{2(1+\bar{t})^2(1+\bar{u}_{\bar{r}})}{[(1+\bar{t})^2+(\bar{u}_{\bar{r}})^2+2\bar{u}_{\bar{r}}]^2}+\frac{2(1+\bar{u}_{\bar{r}})}{(1+\bar{t})^2+(\bar{u}_{\bar{r}})^2+2\bar{u}_{\bar{r}}} -\frac{2}{(1+\bar{u}_{\bar{r}})} \right.  \\ \left. -\frac{2}{(1+\bar{u}_{\bar{r}})^3}\right\} 
 \frac{\rd\bar{u}_{\bar{r}}}{\rd\bar{z}} = \bar{t} \gamma [(3+1/n)\bar{q}]^n(1+\bar{u}_{\bar{r}})^{-(1+3n)},
\label{eq:non_dim_flow_rate_coupled_thick}
\end{multline}
subject to the BC that $\bar{u}_{\bar{r}}(1)= 0$, as before. Unlike, Eq.~\eqref{eq:deflection_ODE_nondim_thin}, Eq.~\eqref{eq:non_dim_flow_rate_coupled_thick} cannot be integrated directly, thus it must be solved numerically. We employ the {\tt odeint} subroutine of the Python package SciPy \cite{SciPy}, with default error tolerances, for this integration.

Equations \eqref{eq:deflection_pressure_thick_nondim} and \eqref{eq:non_dim_flow_rate_coupled_thick} fully specify the static FSI response of the thick-walled hyperelastic cylinder due to the flow of the generalized Newtonian fluid within. Together these two equations can be used to develop the flow rate--pressure relationship for a thick-walled hyperelastic tube, however the calculation must be done via numerical quadratures, unlike the case of the thin-walled cylinder (Sec.~\ref{sec:coupling_thin}).

\section{Results and discussion}
\label{sec:result}

Let us now illustrate the deformation--pressure and flow rate--pressure drop relationships predicted by our FSI theory for the interaction between the steady flow of a power-law fluid within a soft hyperelastic cylindrical vessel containing it. Specifically, in this section, we wish to highlight the effect of hyperelasticity on the structural response of the microtube.

\begin{figure}[t]
\centering
\subfigure[$n=1$ (Newtonian)]{\includegraphics[width=0.5\linewidth]{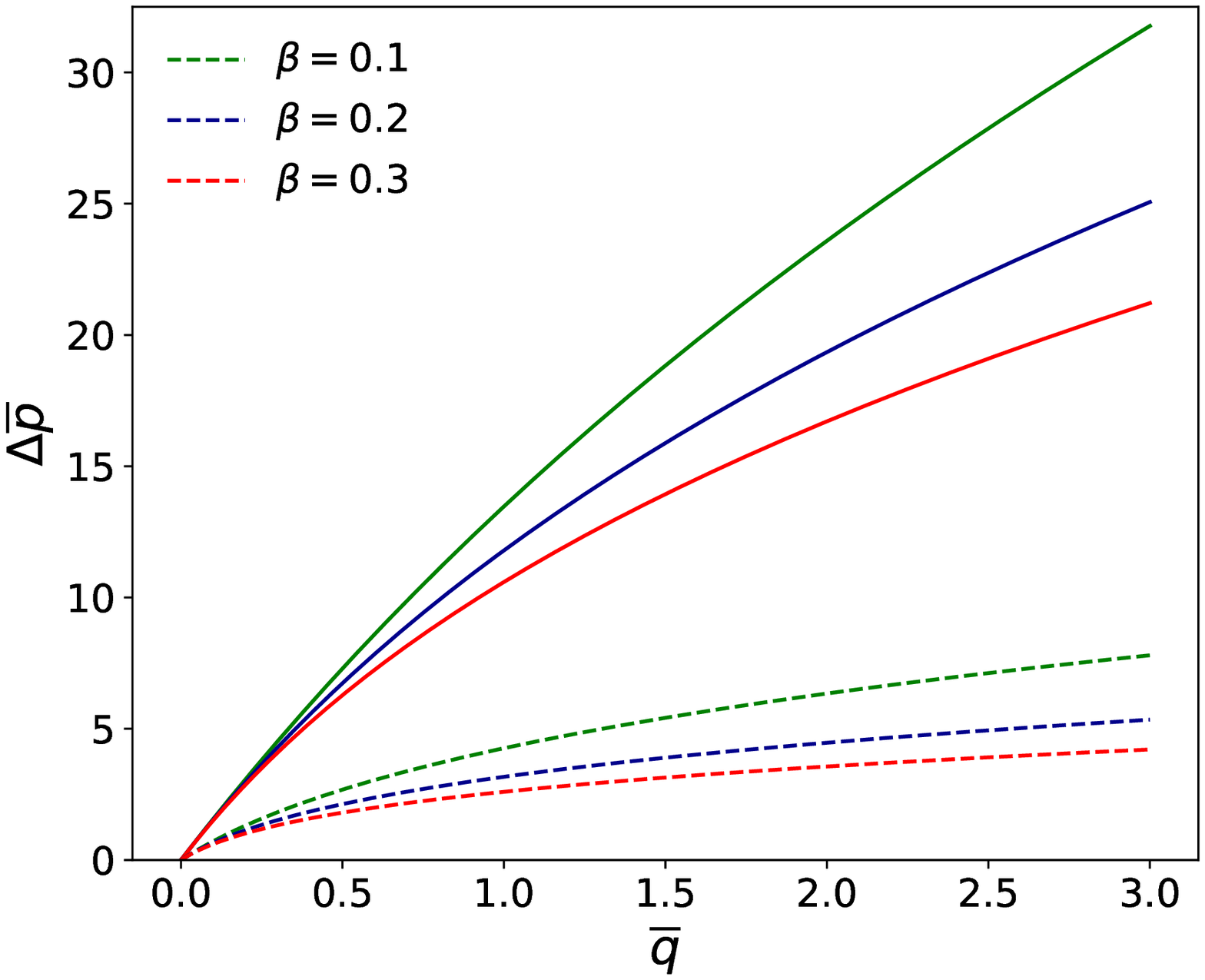}}\hfill
\subfigure[$n=0.7$ (pseudoplastic)]{\includegraphics[width=0.5\linewidth]{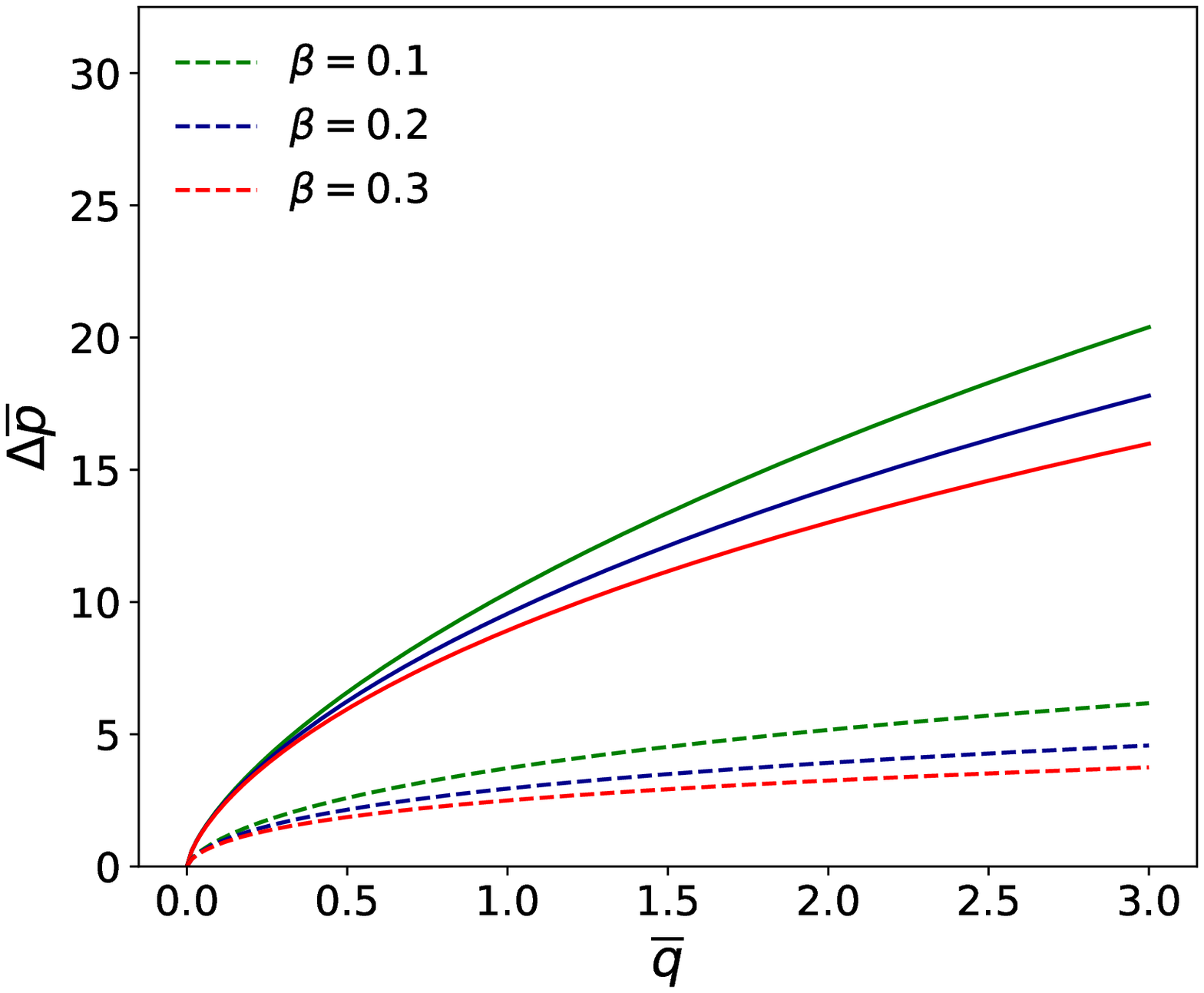}}
\caption{The dimensionless (full) pressure drop across a thin-walled microtube as a function of the dimensionless inlet flow rate $\bar{q}$ for different values of FSI parameter $\beta =3\gamma$. Hyperelastic thin-walled tubes correspond to solid curves, while  linearly elastic thin-walled tubes correspond to dashed curves for (a) a Newtonian fluid with $n=1.0$ (e.g., blood plasma), and (b) a shear-thinning fluid with $n=0.7$ (e.g., whole blood). For the hyperelastic tube, the corresponding theory has been presented in Sec.~\ref{sec:coupling_thin}, while for linearly elastic tubes the predicted pressure drop is given by Eq.~\eqref{eq:q-dp-linear-elastic} derived in \cite{AC18b}.}
\label{fig:Non-dimensional-pressure-thin}
\end{figure}

In Fig.~\ref{fig:Non-dimensional-pressure-thin}, we plot the dimensionless pressure drop $\Delta\bar{p}$ across a thin-walled  microtube as a function of the dimensionless inlet flow rate $\bar{q}$ for different values of the FSI parameter $\beta (=3\gamma)$. The curves (solid) for the thin-walled hyperelastic tube are obtained from the present theory, namely Eqs.~\eqref{eq:pressure_deformation_dimless_thin} and \eqref{eq:flowrate_deflection_nondim}, while the curves (dashed) pertaining to the thin-walled linearly elastic tube are calculated based on the results from our previous study \cite{AC18b}, namely:
\begin{equation}
    \Delta\bar{p} = \frac{1}{(1-\nu/2)\beta}\left(\left\{ 1+2(2+3n){(1-\nu/2)}\beta[(3+1/n)\bar{q}]^n\right\}^{1/(2+3n)} - 1\right),
\label{eq:q-dp-linear-elastic}
\end{equation}
We note that, for both linearly elastic and hyperelastic tubes, the pressure drop decreases with $\beta (=3\gamma)$. This  observation is attributed to the very definition of $\beta$ as the parameter symbolizing the strength of the FSI coupling. For large $\beta$ values, there is ``stronger'' FSI coupling than at small $\beta$ values and, hence, there is larger deformation of the tube. Consequently, the cross-sectional area increases,  lowers the resistance to the flow, and culminates in a smaller pressure drop for large $\beta$ compared to small $\beta$.

Perhaps more attuned to the thesis of this chapter is the difference in the response of a hyperelastic and the responce of a linearly elastic tube for the \emph{same} FSI coupling strength. Thus, comparing the solid and dashed curves, respectively, in Fig.~\ref{fig:Non-dimensional-pressure-thin} at fixed $\gamma$ (i.e., same color), we observe that a hyperelastic tube supports a higher pressure drop than a linearly elastic tube.  We explain this trend by noting that a hyperelastic tube, in general, is stiffer and has a higher tendency to preserve its original configuration compared to a linearly elastic tube. In other words, a hyperelastic material requires higher pressure than a linearly elastic material to sustain the same deformations (see also \cite{LT08a,LT08}).

\begin{figure}[t]
\centering
\subfigure[$n=1$ (Newtonian)]{\includegraphics[width=0.5\linewidth]{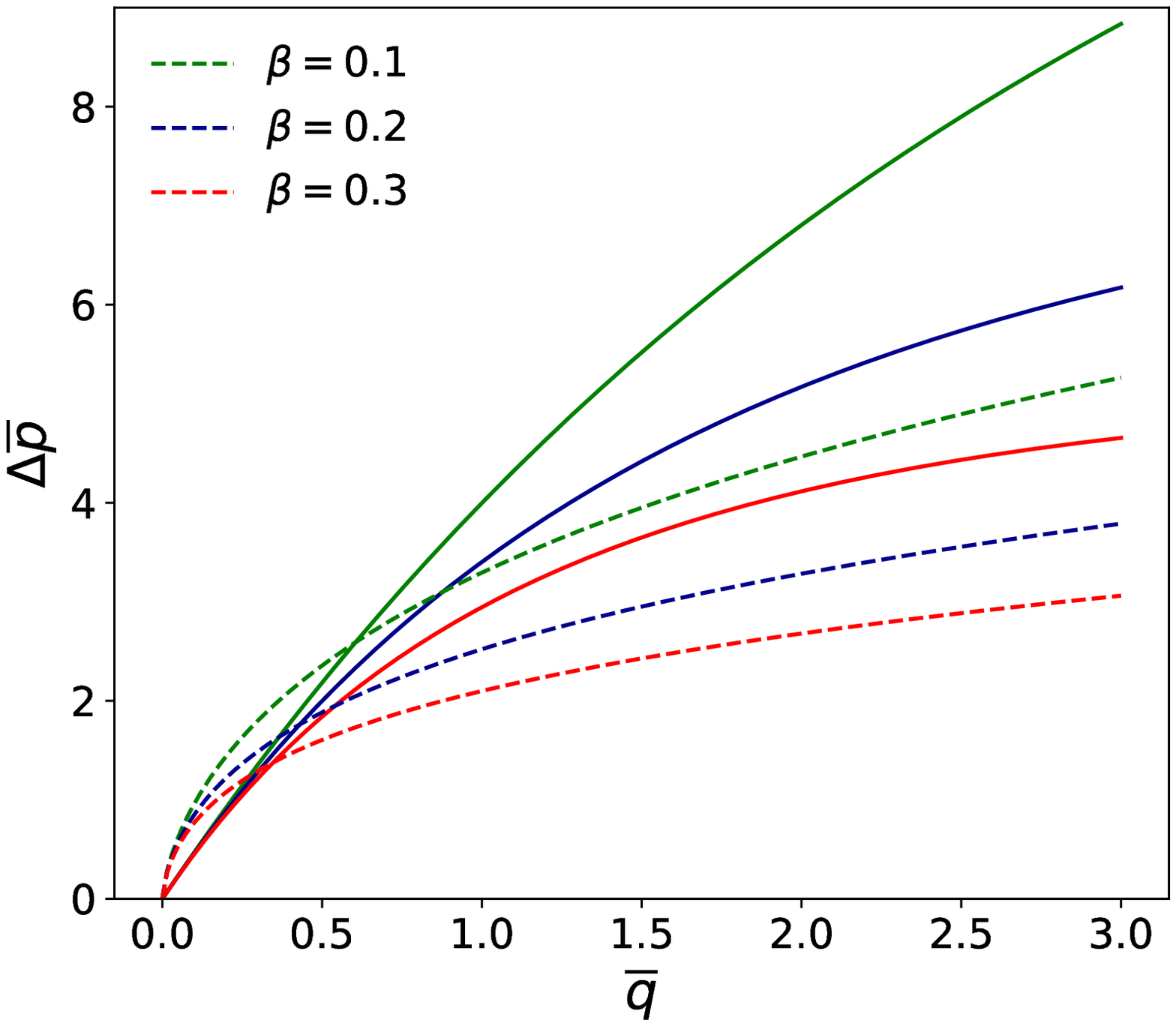}}\hfill
\subfigure[$n=0.7$ (pseudoplastic)]{\includegraphics[width=0.5\linewidth]{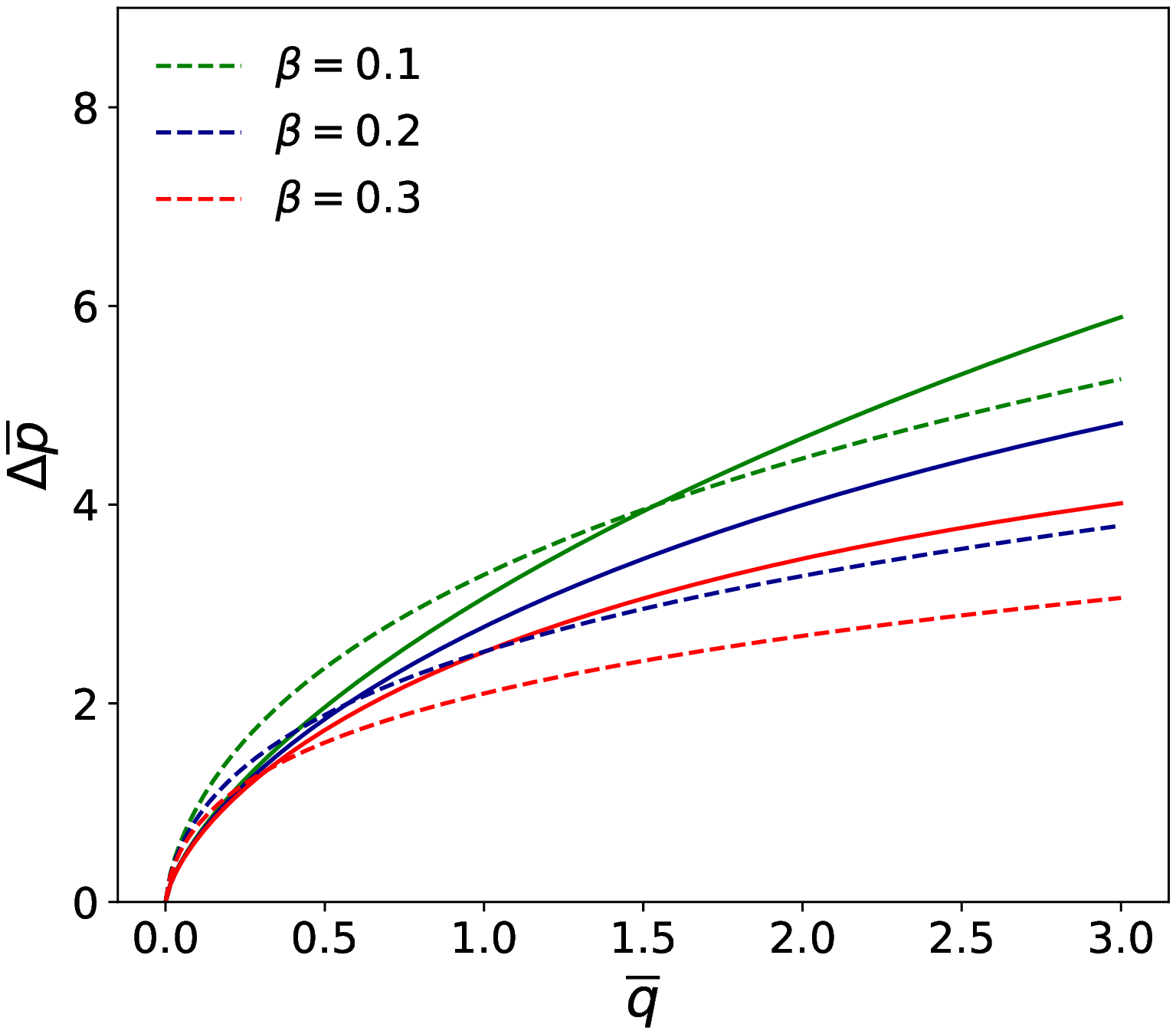}}
\caption{The dimensionless (full) pressure drop across a thick-walled microtube as a function of the dimensionless inlet flow rate $\bar{q}$ for different values of FSI parameter $\beta =3\gamma$ and $\bar{t}=t/a=0.3$. Hyperelastic thick-walled tubes correspond to solid curves, while linearly elastic thick-walled tubes correspond to dashed curves, for (a) a Newtonian fluid with $n=1.0$ (e.g., blood plasma), and (b) a shear-thinning fluid with $n=0.7$ (e.g., whole blood). For a thick-walled hyperelastic tube, the corresponding theory was presented in Sec.~\ref{sec:coupling_thick}, while for thick-walled linearly elastic tubes the predicted pressure drop is given by Eq.~\eqref{eq:q-dp-linear-elastic-thick} derived from the appendix.}
\label{fig:Non-dimensional-pressure-thick}
\end{figure}

Next, we move on to the case of thick-walled tubes and compare, in Fig.~\ref{fig:Non-dimensional-pressure-thick}, the flow rate--pressure drop relation in a thick-walled hyperelastic tube obtained from the present theory, namely Eqs.~\eqref{eq:deflection_pressure_thick_nondim} and \eqref{eq:non_dim_flow_rate_coupled_thick}, with the corresponding relationship for a thick-walled linearly elastic tube calculated based on
\begin{equation}
    \Delta\bar{p} = \frac{1}{\mathfrak{K}\beta} \left( \left\{ 1 + 2(2+3n) 
\mathfrak{K}\beta[(3+1/n)\bar{q}]^n \right\}^{1/(2+3n)} - 1\right),
\label{eq:q-dp-linear-elastic-thick}
\end{equation}
which was derived from Eq.~\eqref{P_vs_Z_Microtube} in the appendix; $\mathfrak{K} = [(1+\bar{t})^2(1+\nu)+(1-2\nu)]/(2+\bar{t})$ and $\bar{t}=t/a$. For the same reason as above, an increase in $\beta (=3\gamma)$ causes $\Delta\bar{p}$ to decrease. 

Things become more interesting, however, when the curves corresponding to same value of the FSI parameter $\beta$ are compared (solid vs.\ dashed) in Fig.~\ref{fig:Non-dimensional-pressure-thick}. For small $\bar{q}$ and small $\beta$, the linearly elastic tube deforms less and sustains a larger $\Delta\bar{p}$ compared to the hyperelastic one. However, this trend is reversed for large $\bar{q}$ and large $\beta$, for which the hyperelastic tube sustains a higher $\Delta\bar{p}$ than the linearly elastic one. This interesting observation can be explained by the very nature of hyperelasticity. It is more difficult to deform a hyperelastic material as the deformation increases, i.e., hyperelastic materials exhibit \emph{strain-stiffening}  \cite{Liu2012,Moon1974,Mihai2011}. Mathematically, this resistance to deformation can be measured through the rate of change of the stress with respect to the strain (or deformation). To this end, when we differentiate the constitutive equation  for a hyperelastic material, namely Eq.~\eqref{eq:constitutive_sigmas}, with respect to the principal deformation along the ``$1$'' direction, whilst keeping the deformations in the other directions constant for simplicity, to  obtain:
\begin{equation}
    \frac{\partial \sigma_1}{\partial \lambda_1}=4\mathbb{C}_1\lambda_1+4\mathbb{C}_2/\lambda_1^3.
\end{equation}
From the latter equation it follows that the resistance to deformation increases with deformation (keeping in mind that $\mathbb{C}_2<0$). On the other hand, for a linearly elastic material, the resistance to deformation is given by Young's modulus $E$ (at least for a uniaxial load), which is a constant!

One could also interpret the much larger $\Delta \bar{p}$ in hyperelastic (over linearly elastic) thin-wall cylinders in Fig.~\ref{fig:Non-dimensional-pressure-thin} as a consequence of strain-hardening. However, in that case, due to significant resistance to deformation of the thin-walled hyperelastic vessel (there is almost one order of magnitude difference in the vertical scales between Figs.~\ref{fig:Non-dimensional-pressure-thin} and \ref{fig:Non-dimensional-pressure-thick}), the interplay between $\bar{q}$ and $\beta$ values just described is not present.

Finally, we also note that an increase in the cylinder thickness will lead to a corresponding decrease in the pressure drop, although this is not shown in Fig.~\ref{fig:Non-dimensional-pressure-thick}, in which all the curves have been plotted for a constant ratio $\bar{t}=t/a=0.3$. This result is similar to the one for thick linearly elastic plates, which was discussed in our previous work \cite{ADC18}, and it can be attributed to an increase in the normal stress throughout the structure's thickness as $\bar{t}$ increases.

\section{Conclusion}
\label{sec:conclusion}

In this chapter, we have solved the problem of steady-state low Reynolds number fluid--structure interaction (FSI) between a generalized Newtonian fluid and a hyperelastic cylindrical tube. The hydrodynamic pressure, which is needed to maintain a unidirectional flow in a deformed cylindrical pipe, was transferred as a load onto the elastic structure, the mechanics of which were analyzed using the \emph{thin} and \emph{thick} pressure vessels theories for isotropic, incompressible Mooney--Rivlin materials. The fluid and solid mechanics were brought together to yield a coupled equation relating the constant inlet flow rate $q$ to the tube's radial deformation $u_r(z)$. For a thin-walled pressure vessel, the latter relation takes the \emph{implicit} dimensional form
\begin{multline}
 \frac{m[(3+1/n){q}]^n}{(\mathbb{C}_1+\mathbb{C}_2)t}(\ell-z) = \frac{[a+{u_r(z)}]^{(3n+2)}}{a^2(3n+2)}+\frac{a^2[a+{u_r(z)}]^{(3n-2)}}{(3n-2)} \\ - \frac{6na^{3n}}{(3n+2)(3n-2)}.
\label{eq:flowrate_deflection_dim}
\end{multline}
As a special case, we have also found the flow rate--deformation relation for a Newtonian fluid ($n=1$, and $m=\mu$ is the shear viscosity):
\begin{equation}
 \frac{4\mu q}{(\mathbb{C}_1+\mathbb{C}_2)t}(\ell-z) = \frac{[a+{u_r(z)}]^{5}}{5a^2}+{a^2[a+{u_r(z)}]} - \frac{6a^{3}}{5},
\label{eq:flowrate_deflection_dim_Newtonian}
\end{equation}
which is an implicit relation for $u_r(z)$ given $q$. Whence, the equation relating the pressure $p(z)$ at an axial location $z$ with the deformation $u_r(z)$ is
\begin{equation}
    p(z) = 2(\mathbb{C}_1+\mathbb{C}_2) \left(\frac{t}{a}\right) \left\{ [1+u_r(z)/a]-\frac{1}{[1+u_r(z)/a]^3} \right\}.
    \label{eq:pressure_deformation_dim_thin}
\end{equation}
Consonant with our previous FSI results \cite{ADC18,AC18b}, the pressure--deformation relationship is set by the structural mechanics alone, hence it \emph{does not explicitly depend} upon the fluid's rheology. 

Equations~\eqref{eq:flowrate_deflection_dim} [or \eqref{eq:flowrate_deflection_dim_Newtonian}] and \eqref{eq:pressure_deformation_dim_thin} fully specify the FSI problem for a thin-walled hyperelastic cylinder. In deriving these equations, we arrived at the dimensionless \emph{FSI parameter} $\gamma$, which determines the ``strength'' of the  coupling between flow and deformation fields:
\begin{equation}
    \gamma := \frac{\mathcal{P}_c}{2(\mathbb{C}_1+\mathbb{C}_2)}\left(\frac{a}{t}\right)
\end{equation}
for a hyperelastic cylinder with material constants $\mathbb{C}_1$ and $\mathbb{C}_2$. Here, the pressure scale $\mathcal{P}_c$ depends on the nature of the physical scenario at hand. For a flow-rate-controlled experiment and/or simulation, as considered in this chapter for example, we set $\mathcal{P}_c = [q/(\pi a^2)]^n m \ell/a^{n+1}$, as dictated by the fluid's momentum balance. On the other hand, for a pressure-drop-controlled experiment or simulation, we can directly set $\mathcal{P}_c = \Delta p$, which means the fluids velocity scale $\mathcal{V}_z$ discussed above becomes a function of the dimensional pressure drop $\Delta p$.

We compared the predicted $q$--$\Delta p$ relation due to the hyperelastic FSI theory developed in this chapter with the corresponding relation due to linearly elastic FSI theory form previous work \cite{AC18b}. In particular, we concluded that a hyperelastic microtube supports smaller deformations than a linearly elastic microtube for the same hydrodynamic pressure, or conversely a hyperelastic microtube can sustain a higher pressure drop than a linearly elastic one, for the same deformation. Finally, our observation in Sec.~\ref{sec:result} that the pressure drop across a soft microtube \emph{decreases}  with the wall thickness is in agreement with the case of rectangular microchannels with a plate for a top wall, which we considered in our previous work \cite{ADC18}. 

In future work, we will report benchmarks of this chapter's purely theoretical considerations against full-fledged three-dimensional FSI simulations and/or experiments (as in \cite{ADC18,AC18b}). Future work could also include extending our approach to FSI between internal viscous flows and composite structures  governed by \emph{generalized} continua continua \cite{AME11}, such as Cosserat continua \cite{MM10} or microstructured materials \cite{DEP18}.

\begin{acknowledgement}
This research was supported by the U.S.\ National Science Foundation under grant No.\ CBET-1705637. We dedicate this work to the 70\textsuperscript{th} anniversary of the director of the Institute of Problems in Mechanical Engineering of the Russian Academy of Sciences: Prof.\ Dr.Sc.\ D.\ I.\ Indeitsev. We also thank Prof. Alexey Porubov for his kind invitation to contribute to this volume, and for his efforts in organizing it.
\end{acknowledgement}

\section*{Appendix}
\addcontentsline{toc}{section}{Appendix}

In this appendix, we consider the flow rate--pressure drop relationship for steady flow of a power-law fluid within a linearly elastic, thick-walled pressure vessel of thickness $t$, and inner radius $r_{i}=a$. The pressure vessel is subject only to an internal distributed pressure load $p$, with zero external pressure. Then, the state of stress evaluated at the inner radius (see \cite{Vable09}) is:
\begin{subequations}\begin{align}
    \sigma_{\theta\theta} &= \left(\frac{r_{o}^2+r_{i}^2}{r_{o}^2-r_{i}^2}\right) p,\\
    \sigma_{rr} &= -p,\\
    \sigma_{zz} &= \left(\frac{r_{i}^2}{r_{o}^2-r_{i}^2}\right) p.
\end{align}\label{eq:stress_linear_thick_wall}\end{subequations}
The hoop strain is given by the constitutive equations of linear elasticity as:
\begin{equation}
    \varepsilon_{\theta\theta} = \frac{u_r}{r_i}=\frac{1}{E}\big[\sigma_{\theta\theta}-\nu(\sigma_{zz}+\sigma_{rr})\big].
    \label{eq:hoop_strain}
\end{equation}
Using Eqs.~\eqref{eq:stress_linear_thick_wall} and \eqref{eq:hoop_strain} yields
\begin{equation}
    \frac{u_r}{r_i}= \frac{1}{E}\left[\left(\frac{r_o^2+r_i^2}{r_o^2-r_i^2}\right) p - \nu\left(\frac{r_i^2}{r_o^2-r_i^2}-1\right)p\right],
\end{equation}
which, upon using Eqs.~\eqref{eq:dimless_vars_structure} and \eqref{eq:Geometry_Inner_Outer}, becomes
\begin{equation}
    \frac{u_r}{r_i} =   \bar{t}\left[\frac{(1+\bar{t})^2(1+\nu)+(1-2\nu)}{(1+\bar{t})^2-1}\right] \beta \bar{p}; \qquad \beta=\frac{\mathcal{P}_c}{E \bar{t}}.
\end{equation}
After deformation, the inner radius is ${R}_{i}=r_i + u_{r}$ (where, initially, $r_i=a$). Thus, the dimensionless inner radius is
\begin{equation}
    \bar{R}_i = \frac{r_i+u_r}{r_i} = 1+\frac{u_r}{r_i}=1+ \left[\frac{(1+\bar{t})^2(1+\nu)+(1-2\nu)}{2 + \bar{t}}\right] \beta \bar{p}.
    \label{eq:deformed_radius_thick_linear}
\end{equation}
Substituting the expression for $\bar{R}_i$ from Eq.~\eqref{eq:deformed_radius_thick_linear} into Eq.~\eqref{eq:dpdz_leading_order}, we obtain an ODE for the dimensionless pressure $\bar{p}$:
\begin{equation}
\frac{\rd \bar{p}}{\rd \bar{z}} = -2[(3+1/n)\bar{q}]^n\left(1+ \mathfrak{K} \beta \bar{p}\right)^{-(1+3n)},
\label{eq:dpdz_leading_order_2}
\end{equation}
where we have defined $\mathfrak{K} := [(1+\bar{t})^2(1+\nu)+(1-2\nu)]/(2+\bar{t})$ for convenience. 
As usual, the ODE~\eqref{eq:dpdz_leading_order_2} is separable and subject to a pressure outlet BC [i.e., $\bar{p}(1)=0$], thus we obtain:
\begin{equation}
\bar{p}(\bar{z}) =  \frac{1}{\mathfrak{K}\beta} \left( \left\{ 1 + 2(2+3n) 
\mathfrak{K}\beta[(3+1/n)\bar{q}]^n (1-\bar{z})\right\}^{1/(2+3n)} - 1\right).
\label{P_vs_Z_Microtube}
\end{equation}
Then, the full pressure drop is simply $\Delta\bar{p}=\bar{p}(\bar{z}=0)$. Note that $\mathfrak{K} = (1-\nu/2) + \mathcal{O}(\bar{t})$, thus the expression for $\Delta\bar{p}$ based on Eq.~\eqref{P_vs_Z_Microtube} [i.e., Eq.~\eqref{eq:q-dp-linear-elastic-thick} above] reduces to Eq.~\eqref{eq:q-dp-linear-elastic} (based on \cite{AC18b}) identically for thin shells ($\bar{t}\ll1$).


\footnotesize{
\bibliographystyle{spbasic.bst}
\bibliography{Mendeley.bib}
}

\end{document}